\documentclass[prb,twocolumn,superscriptaddress]{revtex4-1}

\usepackage{amsmath}
\usepackage{amssymb}
\usepackage{graphicx}

\begin{document}

\title{Field tunable spin density wave phases in Sr$_3$Ru$_2$O$_7$}

\author{C. Lester}
\affiliation{H. H. Wills Physics Laboratory, University of Bristol, Bristol, BS8 1TL, United Kingdom.}

\author{S. Ramos}
\affiliation{School of Physical Sciences, University of Kent, Canterbury, CT2 7NH, United Kingdom.}

\author{R. S. Perry}
\affiliation{London Centre for Nanotechnology and Department of Physics and Astronomy,
University College London, London WC1E 6BT, United Kingdom}

\author{T. P. Croft}
\affiliation{H. H. Wills Physics Laboratory, University of Bristol, Bristol, BS8 1TL, United Kingdom.}

\author{R.~I.~Bewley}
\affiliation{ ISIS Facility, Rutherford Appleton Laboratory, Chilton, Didcot, OX11 0QX, United Kingdom.}

\author{T.~Guidi}
\affiliation{ ISIS Facility, Rutherford Appleton Laboratory, Chilton, Didcot, OX11 0QX, United Kingdom.}

\author{P.~Manuel}
\affiliation{ ISIS Facility, Rutherford Appleton Laboratory, Chilton, Didcot, OX11 0QX, United Kingdom.}

\author{D. D. Khalyavin}
\affiliation{ ISIS Facility, Rutherford Appleton Laboratory, Chilton, Didcot, OX11 0QX, United Kingdom.}

\author{E. M. Forgan}
\affiliation{School of Physics and Astronomy, University of Birmingham, Birmingham B15 2TT, United Kingdom.}

\author{S. M. Hayden}
\affiliation{H. H. Wills Physics Laboratory, University of Bristol, Bristol, BS8 1TL, United Kingdom.}

\maketitle

\textbf{The conduction electrons in a metal experience competing interactions with each other and the atomic nuclei. This competition can lead to many types of magnetic order in metals \cite{White2007_Whit}. For example, in chromium \cite{Fawcett1988_Fawc} the electrons order to form a spin-density-wave (SDW) antiferromagnetic state. A magnetic field may be used to perturb or tune materials with delicately balanced electronic interactions. Here we show that the application of a magnetic field can induce SDW magnetic order in a metal, where none exists in the absence of the field. We use magnetic neutron scattering to show that the application of a large ($\approx 8$~T) magnetic field to the metamagnetic perovskite metal Sr$_3$Ru$_2$O$_7$ (refs \onlinecite{Ikeda2000_IMNK,Perry2001_PGGC,Grigera2004_GGBW,Borzi2007_BGFP,Rost2009_RPMM}) can be used to tune the material through two magnetically-ordered SDW states. The ordered states exist over relatively small ranges in field ($\lesssim 0.4$~T) suggesting that their origin is due to a new mechanism related to the electronic fine structure near the Fermi energy, possibly combined with the stabilising effect of magnetic fluctuations \cite{Berridge2009_BGGS,Berridge2010_BGSG}.  The magnetic field direction is shown to control the SDW domain populations which naturally explains the strong resistivity anisotropy or ``electronic nematic'' behaviour observed \cite{Grigera2004_GGBW,Borzi2007_BGFP} in this material.
 }

\begin{figure*}[ht]
\begin{center}
\includegraphics[width=0.7\linewidth]{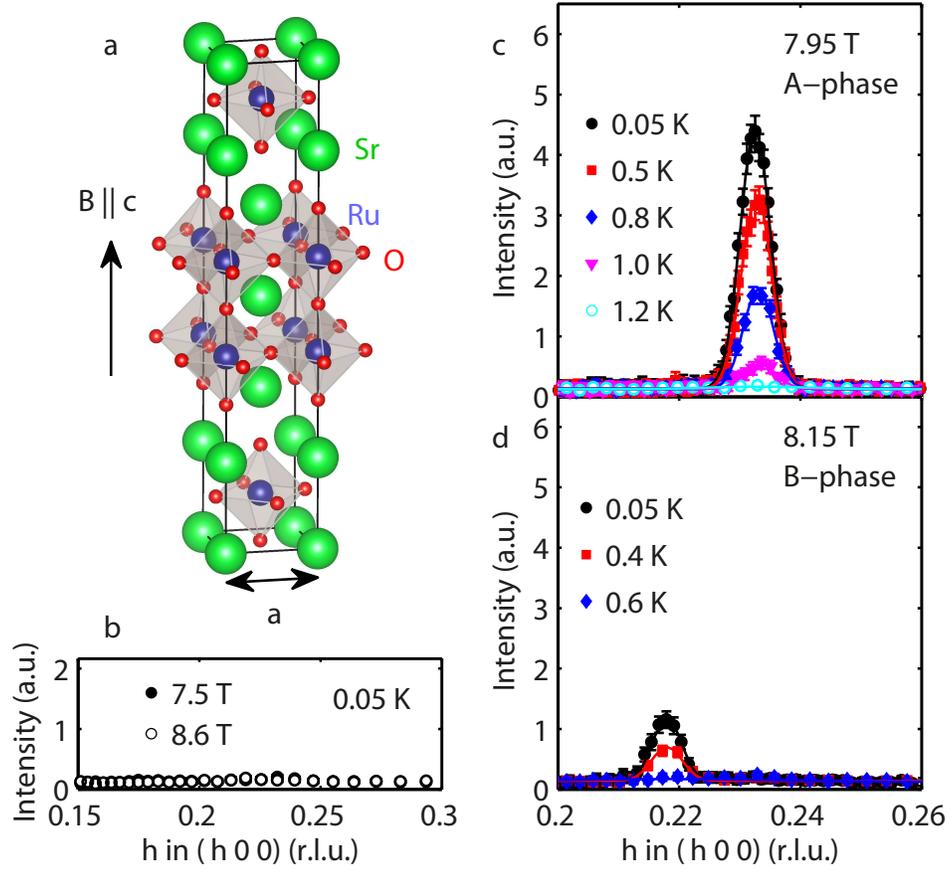}
\end{center}
\caption{\textbf{Spin-density-wave order in Sr$_3$Ru$_2$O$_7$ observed by neutron diffraction.} \textbf{a}, Tetragonal unit cell \cite{Shaked2000_SJCI} of Sr$_3$Ru$_2$O$_7$  showing the direction of applied field. We label reciprocal space, $(h,k,\ell)$ in units of $(2 \pi/a,2 \pi/b,2 \pi/c)$,  based on this cell with $a=3.89$ and $c=20.7$~\AA.\textbf{b}-\textbf{d}, Magnetic Bragg scattering data collected on the WISH spectrometer at various magnetic fields. SDW Bragg peaks are seen at $B$=7.95~T (A-phase, \textbf{c}) and 8.15~T (B-phase, \textbf{d}) with onset temperatures of $1 \pm 0.05$ and $0.5 \pm 0.05$~K respectively. The SDW wavevector $\mathbf{q}_{\mathrm{SDW}}$ changes discontinuously between the two phases. No SDW peaks are seen at $B$=7.5 or 8.6~T (\textbf{b}). }
\label{fig:WISH_Tdep}
\end{figure*}

An itinerant metamagnet \cite{Wohlfarth1962_WoRh,Shimizu1982_Shim} is a metal which undergoes a sudden change from a low- to high- magnetisation state as a function of magnetic field.  We investigated the layered perovskite Sr$_3$Ru$_2$O$_7$ (Fig.~\ref{fig:WISH_Tdep}a). In this material, conduction takes place in the RuO$_2$ bilayers.  Under a magnetic field applied parallel to the $\mathbf{c}$-axis and temperatures below $T \approx 1$~K, Sr$_3$Ru$_2$O$_7$ shows a rapid increase of magnetisation \cite{Rost2009_RPMM} (See Supplementary Fig.~2c) from 0.2 to 0.35~$\mu_B$~Ru$^{-1}$ over a field range of about 1~T near the metamagnetic field, $B_c \approx 7.95$~T.  The metamagnetic behaviour is believed to be caused \cite{Wohlfarth1962_WoRh,Shimizu1982_Shim,Millis2002_MSLG,Binz2004_BiSi} by proximity to ferromagnetism and the band structure having a local minimum in the density of states at the Fermi energy ($\varepsilon_F$) and/or a maximum  near  $\varepsilon_F$. Such features may result from a van-Hove singularity near $\varepsilon_F$, as observed \cite{Tamai2008_TAMM} by angle-resolved photo-emission spectroscopy.

\begin{figure*}[ht]
\begin{center}
\includegraphics[width=0.95\linewidth]{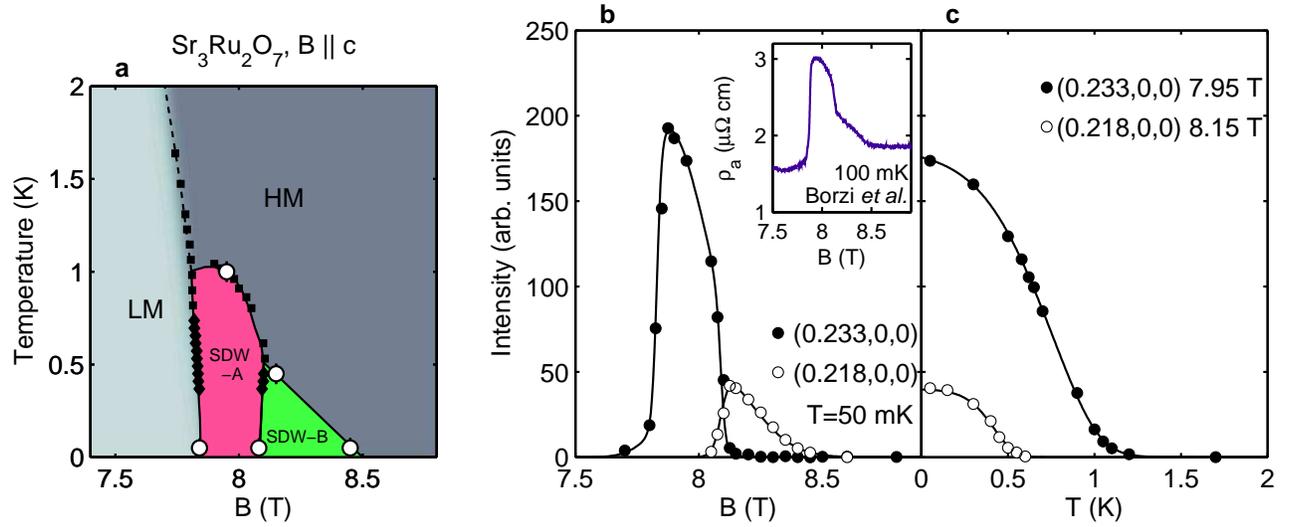}
\end{center}
\caption{\textbf{Magnetic phase diagram of Sr$_3$Ru$_2$O$_7$.} \textbf{a}, Phase diagram of Sr$_3$Ru$_2$O$_7$ determined from the present neutron scattering data combined with magnetic and transport measurements \cite{Grigera2004_GGBW,Borzi2007_BGFP}. Dotted line represents the peak in $\chi$ separating low-moment (LM) and high-moment (HM) phases. Solid lines defining the A-phase boundaries are determined from maxima in $\chi$, $d\rho/dB$, and $d^2\rho/dT^2$ (refs~\onlinecite{Grigera2004_GGBW,Borzi2007_BGFP}) and reproduced (see Supplementary Fig.~2) on the present samples (diamonds and squares).  SDW-A and SDW-B are spin-density-wave phases with different wavevectors. Open circles are phase boundaries determined from the $B$ and $T$ dependence of Bragg peak intensities. The high temperature boundary of SDW-B (solid line) is assumed to vary linearly with $B$. \textbf{b},\textbf{c}, Magnetic field and temperature dependencies of the $\mathbf{q}_{SDW}^A$ and $\mathbf{q}_{SDW}^B$ Bragg peak intensities. The inset to \textbf{b} shows the field-dependent resistivity $\rho_{a}(B)$ for a field parallel to $\mathbf{c}$ (Ref. \onlinecite{Borzi2007_BGFP}.)}
\label{fig:WISH_phase_diagram}
\end{figure*}

\begin{figure*}[ht]
\begin{center}
\includegraphics[width=0.95\linewidth]{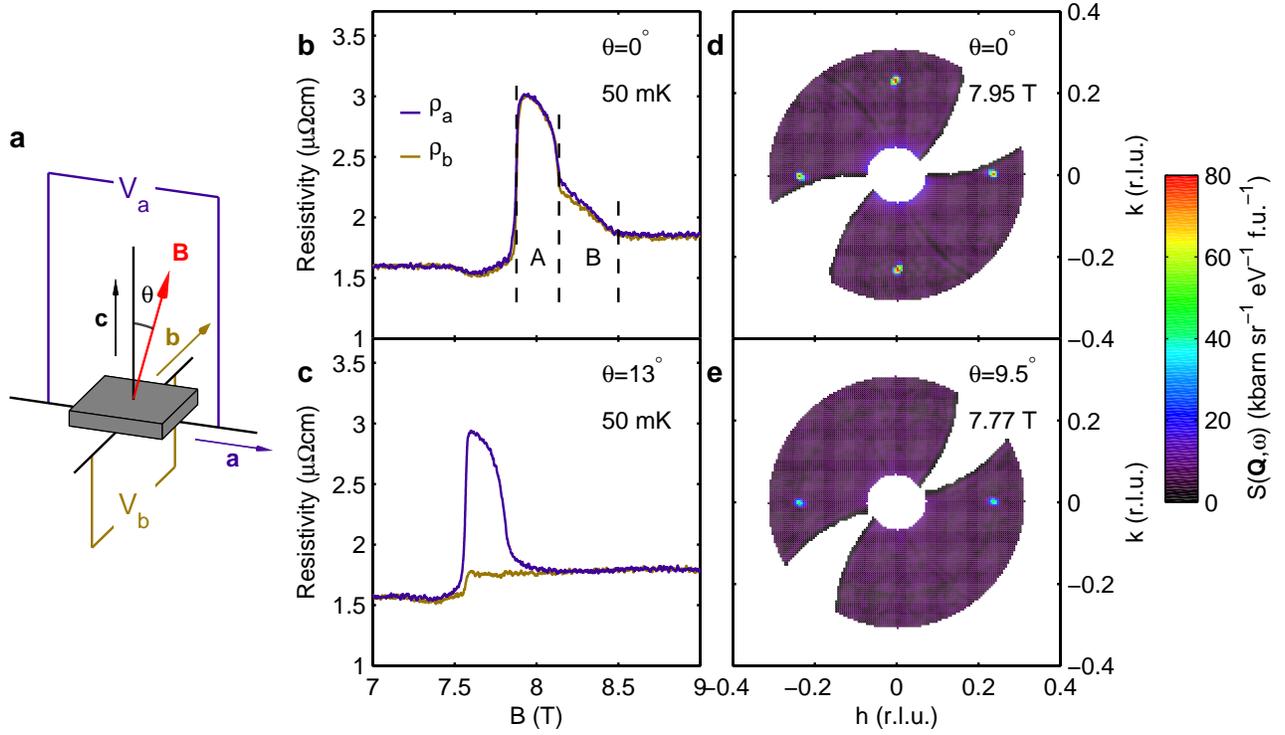}
\end{center}
\caption{\textbf{The effect of tilting the magnetic field away from the c-axis on the SDW order and magnetoresistance.} \textbf{a}-\textbf{c}, Anisotropic resistivity induced by tilted fields after Borzi \textit{et al.} \cite{Borzi2007_BGFP}. For a magnetic field parallel to the crystalline $c$-axis, $\rho_a$ and $\rho_b$ are almost identical. An in-plane component of the field along the $a$-axis is created by tilting the magnetic field away from the $c$-axis and causes a dramatic reduction in $\rho_b$. \textbf{d},\textbf{e}, Elastic (Bragg) scattering for $T=50$~mK. Data are integrated over $-0.1<l<0.1$~r.l.u. and $-15<E<15$~$\mu$eV. For fields parallel to the $c$-axis (\textbf{d}), four Bragg satellites are seen corresponding to two SDW domains with wavevectors parallel to $\mathbf{a}^{\star}$ and $\mathbf{b}^{\star}$. Tilting the field to give a component along the $a$-axis suppresses the SDW domain propagating along $\mathbf{b}^{\star}$.  The large increase in resistivity (\textbf{b},\textbf{c}) is caused by a SDW with a propagation vector parallel to the current flow.}
\label{fig:LET_resistivity}
\end{figure*}

A unique feature of Sr$_3$Ru$_2$O$_7$ is the observation of an unusual phase (denoted as ``A'', see Fig.~\ref{fig:WISH_phase_diagram}a) near $B_c$ for $\mathbf{B} \parallel \mathbf{c}$ and for $T \lesssim 1$~K. The A-phase is a region of higher resistivity (see Fig.~\ref{fig:LET_resistivity}b) whose boundaries can be identified from anomalies in a.c. susceptibility \cite{Grigera2004_GGBW}, resistivity \cite{Grigera2004_GGBW}, NMR (ref.~\onlinecite{Kitagawa2005_KIPT}) and magnetostriction \cite{Stingl2011_SPMG}. Tilting the field $\mathbf{B}$ away from $\mathbf{c}$ to give a component along $\mathbf{a}$ or $\mathbf{b}$ induces a large anisotropy (``electron nematic'' behaviour) in the in-plane resistivity both in the A-phase \cite{Borzi2007_BGFP} and the adjacent region \cite{Bruin2013_BBGR} in the $B-T$ plane.  For example, a $\mathbf{B}$ component along $\mathbf{a}$ causes $\mathbf{b}$ to become the easy direction for current flow (see Fig.~\ref{fig:LET_resistivity}a-c).

\begin{figure*}[ht]
\begin{center}
\includegraphics[width=0.6\linewidth]{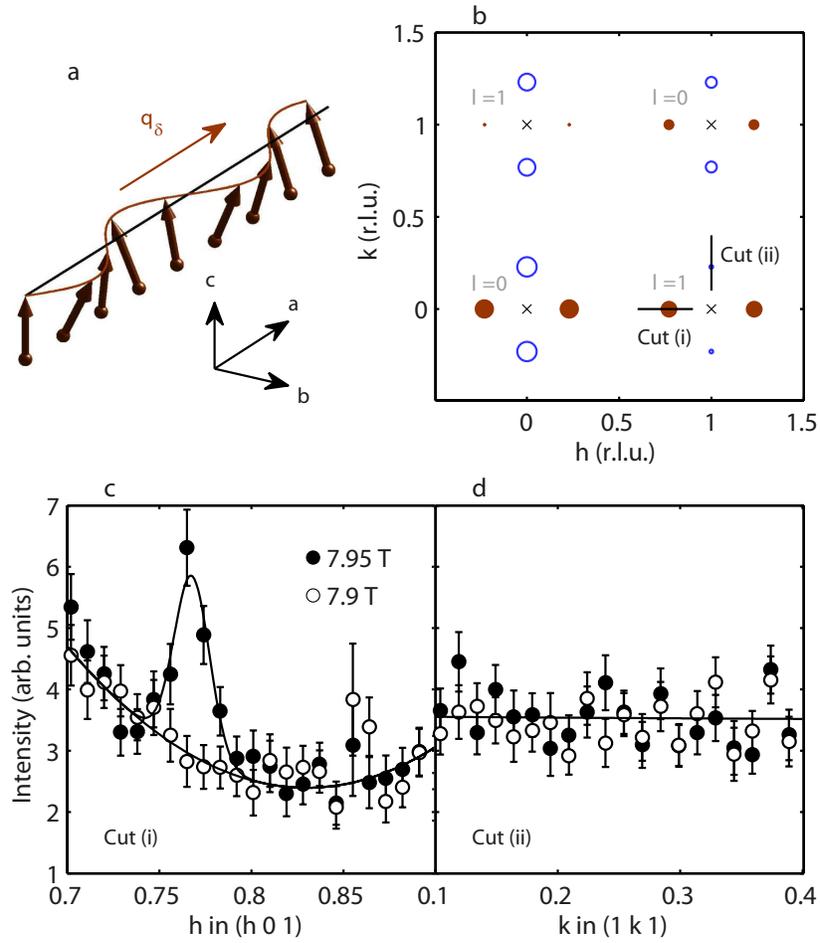}
\end{center}
\caption{ \textbf{The structure of the SDW.} \textbf{a} Proposed structure of a single domain of the SDW A-phase ($\mathbf{B} \parallel  \textbf{c}$) in Sr$_3$Ru$_2$O$_7$, a transverse spin density wave with $\mathbf{q}_{\mathrm{SDW}}=(\delta,0,0)$. Arrows represent moments on the Ru sites of Fig.~\ref{fig:WISH_Tdep}a. \textbf{b}, The SDW domain in \textbf{a} gives rise to satellite peaks (closed circles) at $\mathbf{Q}=\boldsymbol{\tau} \pm \mathbf{q}_{\mathrm{SDW}}$.  Open circles are peaks associated with the domain with $\mathbf{q}_{\mathrm{SDW}}=(0,\delta,0)$. The areas of the circles represent the intensities of the satellite Bragg peaks for the structure in \textbf{a}.   \textbf{c}, Elastic data collected using LET along cut (i) in \textbf{b} showing a SDW Bragg peak. \textbf{d}, Data collected along cut (ii) in \textbf{b}  show SDW peak with zero or small intensity consistent with the structure in \textbf{a}.}
\label{fig:LET_spin_structure}
\end{figure*}

Motivated by previous reports \cite{Capogna2003_CFHW,Ramos2008_RFBH} of strong low-energy spin fluctuations, we searched for static SDW order with higher energy resolution and lower temperatures.
In SDW antiferromagnets the ordered moment is modulated in space with a wavevector $\mathbf{q}_\mathrm{SDW}$. This results in satellite peaks at reciprocal space positions $\mathbf{Q}=\boldsymbol{\tau}+\mathbf{q}_{\mathrm{SDW}}$, where $\boldsymbol{\tau}$ is a reciprocal lattice point (including $\boldsymbol{\tau}=0$) of the crystal structure. Fig.~\ref{fig:WISH_Tdep}c shows Bragg scans along $\mathbf{Q}=(h,0,0)$ for a magnetic field $B_c$=7.95~T and a series of temperatures in the range $0.05 < T < 1.2$~K which traverse the A-phase (see Fig.~\ref{fig:WISH_phase_diagram}a).  Below  $T \approx 1.0$~K a magnetic Bragg peak develops at $\mathbf{q}_{\mathrm{SDW}}^{A}=(0.233 \pm 0.002,0,0)$. Scans along other directions parallel to $\mathbf{b}^{\star}$ and $\mathbf{c}^{\star}$ (see Supplementary Fig. 4) show that the peak is sharp in all three directions indicating 3D magnetic ordering with in-plane correlation lengths greater than 350~\AA. Energy-dependent scans through the ordering position (see Supplementary Fig. 3) show that the peak is resolution limited in energy. This implies that the inverse lifetime of any magnetic fluctuations is less than $\tau^{-1}=$4~$\mu$eV$\approx$40~mK$\approx$1~GHz.  From the intensity of the Bragg peak we estimate the magnitude of the ordered moment (for $T=50$~mK and $B=7.95$~T) to be $\langle m_q \rangle =0.10 \pm 0.02$~$\mu_B$~Ru$^{-1}$ assuming the structure in Fig.~\ref{fig:LET_spin_structure}a.  Fig.~\ref{fig:WISH_phase_diagram}b shows the intensity of the $\mathbf{q}_{SDW}^{A}$ Bragg peak measured as a function of magnetic field. We find that the onset of the Bragg peak coincides with the boundaries of the A-phase indicating that it is associated with SDW order.

The susceptibility and resistivity (see Fig.~\ref{fig:LET_resistivity}b) of Sr$_3$Ru$_2$O$_7$ show anomalous behaviour for fields above the A-phase boundary \cite{Borzi2007_BGFP,Bruin2013_BBGR,Grigera2004_GGBW}. We therefore also searched for SDW order in this region.  For fields greater than the upper boundary of the A-phase, we observe (see Fig.~\ref{fig:WISH_Tdep}d) an incommensurate peak at a different wavevector $\mathbf{q}_{\mathrm{SDW}}^{B}=(0.218 \pm 0.002,0,0)$ to that observed in the A-phase. We denote this second SDW-ordered region the ``B-phase''.

The temperature and field dependencies of the $\mathbf{q}_{\mathrm{SDW}}^{A}$ and $\mathbf{q}_{\mathrm{SDW}}^{B}$ Bragg peak intensities are shown in Fig.~\ref{fig:WISH_phase_diagram}b,c. For the $\mathbf{q}_{\mathrm{SDW}}^{A}$ Bragg peak, we find that the fields and temperatures at which the magnetic order disappears coincide closely with the boundaries of the A-phase determined from a.c. susceptibility and resistivity \cite{Grigera2004_GGBW}. The boundaries of the B-phase for $\mathbf{B} \parallel \mathbf{c}$ are less well characterised. Borzi \textit{et al}. \cite{Borzi2007_BGFP} observe a high field tail to the resistivity anomaly for $8.1 \lesssim B \lesssim 8.5$~T and $T=50$~mK, which defines the width of the B-phase in field. For a small tilt of the magnetic field away from the $c$-axis, Bruin \textit{et al}. \cite{Bruin2013_BBGR} identify a region of anisotropic resistance which persists up to 0.4~K and appears to correspond to the B-phase.

The SDW modulation of the ordered moment [$\mathbf{m(r)}$] results in satellite peaks. For the bct lattice of Sr$_3$Ru$_2$O$_7$, we expect SDW peaks around the $\boldsymbol{\tau}=$(0,0,0), (1,0,1), (0,1,1) and (1,1,0) reciprocal lattice points (Fig.~\ref{fig:LET_spin_structure}b). We observe satellite peaks along $(h,0,1)$, but not along $(0,k,1)$. The simplest structure consistent with this observation is the linear transverse SDW shown in Fig.~\ref{fig:LET_spin_structure}a. Other structures such as a cycloid \cite{Berridge2009_BGGS,Berridge2010_BGSG} with $\mathbf{m(r)}$ in the $ab$ plane or a modulation of the moment $\mathbf{m(r)}$ parallel to $\mathbf{c}$ would give satellite peaks along $(0,k,1)$ with similar intensity to those along $(h,0,1)$.

One of the most fascinating properties of the A-phase is its electron nematic behaviour \cite{Borzi2007_BGFP}.  For magnetic fields parallel to the $c$-axis, the A-phase is associated with a dramatic increase in the resistivity $\rho$ which is isotropic with respect to the direction of charge transport within the RuO$_2$ planes (see Fig.~\ref{fig:LET_resistivity}b). By tilting the magnetic field away from the $c$-axis we can introduce an in-plane (IP) component of magnetic field $\mathbf{B}_{\mathrm{IP}}$ along the $a$-axis.  Under these conditions, charge transport in the A-phase becomes strongly anisotropic. For current parallel to the in-plane field and the $a$-axis, the resistance anomaly in $\rho_a(B)$ associated with the A-phase remains (Fig.~\ref{fig:LET_resistivity}b), while for currents perpendicular to $\mathbf{B}_{\mathrm{IP}}$, the anomaly in $\rho_b(B)$ is completely suppressed (Fig.~\ref{fig:LET_resistivity}c).

We investigated the effect of similar tilted fields on the SDW order of the A-phase. For fields parallel to the $c$-axis (see Fig.~\ref{fig:LET_resistivity}d) we observe two domains of the transverse SDW, one with a propagation vector $\mathbf{q}$ parallel to $\mathbf{a}^{\star}$ giving peaks at $(\pm 0.233,0,0)$ and the other with $\mathbf{q}$ parallel to $\mathbf{b}^{\star}$ giving peaks at $(0,\pm 0.233,0)$. When the field is tilted to give a component parallel to $\mathbf{a}$, the domain propagating along $\mathbf{b}$ is completely suppressed (Fig.~\ref{fig:LET_resistivity}e).  Thus, the applied magnetic field allows fine control of domain populations and the presence of the SDW domain modulated along $\mathbf{a}$ (Fig.~\ref{fig:LET_resistivity}e) is associated with the resistance anomaly in $\rho_a$ (Fig.~\ref{fig:LET_resistivity}c).

The existence of a SDW provides a natural explanation for the resistivity anomalies observed in Sr$_3$Ru$_2$O$_7$.  SDW order in metals may increase the resistivity by gapping out the Fermi surface or by introducing additional scattering mechanisms due to the excitations associated with the magnetic order \cite{Fawcett1988_Fawc}. For Sr$_3$Ru$_2$O$_7$, the field-dependent resistivity shown in Fig.~\ref{fig:LET_resistivity}b closely tracks the sum of the two magnetic order parameters (as measured by the SDW Bragg peak intensities) for the A and B phases (Fig.~\ref{fig:WISH_phase_diagram}b). The gapping of the Fermi surface will be closely related to the SDW order parameter, hence we believe the removal of electronic states from near the Fermi surface is the most likely cause of the resistivity anomalies.

There are other examples of inhomogeneous magnetically modulated states induced by a magnetic field. For example, the application of a magnetic field to certain insulating quantum magnets such as TlCuCl$_3$ causes a Bose-Einstein condensation of magnons and antiferromagnetic order \cite{Zapf2014_ZaJB}.  In the heavy fermion superconductor CeCoIn$_5$ (ref.~\onlinecite{Kenzelmann2008_KSNS}), a ``Q-phase'' with spatially modulated superconducting and magnetic order parameters is created abutting $B_{c2}$. Since Sr$_3$Ru$_2$O$_7$ is neither a superconducting metal nor an insulator, the field-induced order must have a different mechanism to these two cases.

The formation of SDWs in metals (e.g. Cr) is usually described in terms of a Stoner theory including a wavevector-dependent-susceptibility $\chi_0(\mathbf{q})$ and an exchange interaction parameter $I$ (ref.~\onlinecite{White2007_Whit,Fawcett1988_Fawc}). SDW order occurs when the generalised Stoner criterion $\chi_0(\mathbf{q}) I \geq 1$ is satisfied. The ordering wavevector $\mathbf{q}_{\mathrm{SDW}}$ is determined from the peak in $\chi_0(\mathbf{q})$ and ultimately by Fermi surface nesting.  In the vicinity of a metamagnetic transition, the Fermi surface changes rapidly with field as the Fermi energy of one of the spin species passes through a peak in the density of states. Such a rapid change may lead to a SDW phase that is only favoured over a narrow range in field \cite{Rice1970_Rice}. In Sr$_3$Ru$_2$O$_7$, two slightly different SDW states can be favoured. We note that the $\alpha_1$ and $\gamma_3$ sheets \cite{Tamai2008_TAMM} provide approximately the right nesting vectors to match  $\mathbf{q}_{\mathrm{SDW}}^{A}$ and $\mathbf{q}_{\mathrm{SDW}}^{B}$ (refs~\onlinecite{Capogna2003_CFHW,Tamai2008_TAMM,Singh2001_SiMa}).  It has recently been proposed \cite{Berridge2009_BGGS,Berridge2010_BGSG}  - based on a microscopic Stoner theory and a Landau-Ginzburg expansion - that transverse spin modulated states can be further stabilized by soft transverse magnetic fluctuations \cite{Wohlfarth1962_WoRh,Shimizu1982_Shim}.

The linear transverse nature of the SDW is not obviously predicted by the Landau-Ginzburg theory \cite{Berridge2009_BGGS,Berridge2010_BGSG}. However, linearly polarised SDWs such as the one observed here (also in metals such as Cr, ref.~\onlinecite{Fawcett1988_Fawc}) may be favoured by additional contributions \cite{Overhauser1962_Over,Walker1980_Walk} to the free energy.
In the absence of strong anisotropy, antiferromagnets usually favour $\mathbf{m}_\mathbf{q} \perp \mathbf{B}$ and hence favour the structure for the A-phase shown in
Fig.~\ref{fig:LET_spin_structure}a. This structure would naturally host single-$\mathbf{q}$ domains which would be enhanced or suppressed by tilting the field away from $\mathbf{c}$, leading to the population imbalance shown in Fig.~\ref{fig:LET_resistivity}e. The imbalance  naturally explains the observed ``nematic'' (anisotropic) transport properties of the SDW-A phase.

One can ask whether other examples of field-induced SDWs can be found in metals and what conditions are required for their existence. The relevant special properties of Sr$_3$Ru$_2$O$_7$ may include its two-dimensional electronic structure, nested Fermi surface, the existence of a van Hove singularity near the Fermi energy and concomitant metamagnetic transition, and its strongly enhanced magnetic susceptibility.

\section*{Acknowledgments}

We acknowledge helpful discussions with A.P. Mackenzie, R. Coldea, Y. Maeno, R. Evans, and R. M. Richardson.  We are grateful to S. A. Grigera and A.P. Mackenzie for providing resistivity data from ref~\onlinecite{Borzi2007_BGFP} which is reproduced in Figs.~\ref{fig:WISH_phase_diagram} and \ref{fig:LET_resistivity}. Our work was supported by the UK EPSRC (Grant No. EP/J015423/1).

\section*{Additional Information}
 Correspondence and requests for materials should be addressed to S.M.H. (s.hayden@bristol.ac.uk).

\newpage

\newpage

\newpage

\newpage

\end{document}